\documentclass{aa} 
\usepackage{txfonts}

\usepackage{graphicx}
\usepackage{hyperref}
\usepackage{natbib}

\newcommand{\Fg}[1]{Figure~\ref{fig:#1}}

\newcommand{\Fgs}[2]{Figures\ \ref{fig:#1} and \ref{fig:#2}}

\newcommand{\eq}[1]{Eq.~(\ref{eq:#1})\xspace}
\newcommand{\Eq}[1]{Equation~(\ref{eq:#1})\xspace}

\newcommand{\Eqss}[3]{Eqs.\ (\ref{eq:#1}), (\ref{eq:#2}) and (\ref{eq:#3})}

\begin{document}

   \title{Growth after the streaming instability: \\
   The radial distance dependence of the planetary growth}

   \author{Hyerin Jang
          \inst{1,2}
          \and
          Beibei Liu\inst{3\star} 
          \and 
          Anders Johansen\inst{4,2}}

   \institute{Department of Astrophysics/IMAPP, Radboud University, P.O. Box 9010, NL-6500 GL Nijmegen, The Netherlands
         \and
         Lund Observatory, Department of Astronomy and Theoretical Physics, Lund University, Box 43, 22100 Lund, Sweden
         \and
          Institute for Astronomy, School of Physics, Zhejiang University, Hangzhou 310027, China
        \and
            Center for Star and Planet Formation, GLOBE Institute, University of Copenhagen, Øster Voldgade 5-7, 1350 Copenhagen, Denmark\\
            \email{hyerin.jang@astro.ru.nl,bbliu@zju.edu.cn,Anders.Johansen@sund.ku.dk}}

   \date{\today}

  \abstract
   {
  Streaming instability is hypothesized to be triggered at particular protoplanetary disk locations where the volume density of the solid particles is enriched comparable to that of the gas. A ring of planetesimals thus forms when this condition is fulfilled locally. These planetesimals collide with each other and accrete inward drifting pebbles from the outer disk to further increase masses. We investigate the growth of the planetesimals that form in a ring-belt at various disk radii. Their initial mass distributions are calculated based on the formula summarized from the streaming instability simulations. We simulate the subsequent dynamical evolution of the planetesimals with a protoplanetary disk model based either on the minimum mass solar nebula (MMSN) or on the Toomre stability criterion. For the MMSN model, both pebble accretion and planetesimal accretion are efficient at a close-in orbit of $0.3$ AU, resulting in the emergence of several super-Earth mass planets after $1$ Myr. For comparison, only the most massive planetesimals undergo substantial mass growth when they are born at $r{=}3$ AU, while the planetesimals at $r{=}30$ AU experience little or no growth. On the other hand, in the denser Toomre disk, the most massive forming planets can reach Earth mass at $t{=}1$ Myr and reach a mass between that of Neptune and that of Saturn within $3$ Myr at $30$ AU and $100$ AU.
  Both the pebble and planetesimal accretion rate decrease with disk radial distance. Nevertheless, planetesimal accretion is less pronounced than pebble accretion at more distant disk regions. Taken together, the planets acquire higher masses when the disk has a higher gas density, a higher pebble flux, and/or a lower Stokes number of pebbles.
   }


   \maketitle
%
\section{Introduction}
\paragraph{}
The growth of millimeter- to centimeter-sized bodies (named pebbles hearafter) into kilometer- or larger sized planetesimals faces a challenge in the classical planet formation theory (see \citet{Johansen2014}, \citet{Liu_Ji2020} and \citet{Drazkowska2022} for reviews). Their growth is mainly limited by bouncing, fragmentation \citep{Guttler_etal2010,Zsom_etal2010}, and fast radial drift \citep{Weidenschilling1977}. The streaming instability mechanism provides a promising solution to overcome the above barriers, when the pebbles self-concentrate together and collapse into planetesimals due to the collective effect of gravity \citep{Youdin_Goodman2005,Johansen2007}.

In order to initiate the streaming instability, the volume density of pebbles needs to be enriched comparable to that of the gas, $\rho_{\rm peb}{\simeq}\rho_{\rm g}$ \citep{Youdin_Goodman2005}. This criterion is also frequently written in terms of the disk metallicity with a surface density ratio, $Z{=}\Sigma_{\rm peb}/\Sigma_{\rm g}$. Literature numerical simulations find that a super-solar metallicity (${\gtrsim}2{-}5\%$) is generally required for the onset of the streaming instability \citep{Carrera_etal2015,Yang_etal2017,Li_etal2021}. Specifically, \cite{Yang_etal2017} found particles of $\tau_s=10^{-2}$ and $10^{-3}$ are concentrated by the streaming instability when the disk metallicities are $1-2\%$ and $3-4\%$, respectively. \cite{Li_Youdin2021} suggested the metallicity could be lower than 1\% with $\tau_s > 0.01$ in their particle clumping simulations. Higher $Z$ might be needed if considering a size distribution of pebbles \citep{Zhu_Yang2021}.  Overall, the threshold metallicity is difficult to satisfy globally in a canonical solar-like protoplanetary disk. 

Various hypotheses have been proposed so that the enhancement of the pebble surface density can be fulfilled locally at peculiar protoplanetary disk locations. For instance, the ice-line location of different chemical species, such as water, silicate, and carbon monoxide \citep{Ros_Johansen2013,Schoonenberg_Ormel2017,Drazkowska_Alibert2017,Hyodo_etal2019} can act as planetesimal nurseries, since large pebbles can grow there by condensation.
Pressure bumps are potential sites for triggering the streaming instability where drifting pebbles get accumulated and the pressure gradient varies.\footnote{ The streaming instability does not formally apply at the pressure maximum point, since the pressure gradient is zero there.} Another potential location is the dead-zone inner edge. Since the ionization level and density are different between the turbulent active region and quiescent dead-zone region, their boundary can naturally trap drifting pebbles and induce streaming instability \citep{Chatterjee_Tan2014,Ueda_Flock2019}. Furthermore, pebbles might become clumped at the vortex generated by hydrodynamical instabilities \citep{Surville_etal2016} or the spiral arms in self-gravitating disks \citep{Gibbons_etal2012,Elbakyan_etal2020}. Taken together, the triggering site can broadly span from the most inner sub-AU disk region to the most outer disk regions extending to a few tens or hundreds of AUs.

When the streaming instability is triggered at a single-site disk location, a ring of planetesimals forms rapidly \citep{Liu_etal2019}. Their birth masses follow a top-heavy distribution, which can be approximately described by a power-law plus exponential decay function \citep{Johansen_etal2015apr,Schafer_etal2017, Abod_etal2019}. The planetesimals have a characteristic mass  of ${\sim }10^{-6} \ M_{\oplus}$ ($100$ km in size) when they form at the asteroid belt region in the minimum mass solar nebula (MMSN) model \citep{Johansen_etal2015apr}. The mass increases with gas surface density, disk metallicity,  and radial distance \citep{Liu_etal2020}.

After their formation, these planetesimals can further increase their masses by colliding with other planetesimals in the birth ring, as well as by accreting inwardly drifting pebbles from the outer disk. As expected, planetesimal accretion \citep{Kokubo_Ida1996} and pebble accretion \citep{Ormel_Klahr2010,Lambrechts_Johansen2012} operate together to contribute to the subsequent mass growth. In this paper, we assume a streaming instability-induced planetesimal formation at specific disk locations. We investigate the subsequent mass growth of these planetesimals in a ring-belt structure by a combination of planetesimal and pebble accretion. The growth at the conventional water-ice-line location ($r_{\text{ice}} {=} 2.7$ AU) has been investigated by \cite{Liu_etal2019}. We follow their numerical construction and mainly focus on how the growth patterns vary as a function of radial distance.

We introduce the model setup in Section \ref{sec:method}. In Section \ref{sec:results}, we show different dynamical evolution of the planetesimals at $0.3$ AU, $3$ AU, and $30$ AU in the MMSN model. We also explore how the model parameters (planet migration,  pebble flux, Stokes number of pebbles) affect the final forming planets in Section \ref{sec:parameters}. We adopt the upper limit of a stable disk under the Toomre's criterion and investigate whether giant planets can form early at the distant orbits of $30$ and $100$  AU in Section \ref{sec:superEarth}. The discussion and final conclusions are summarized in Sections \ref{sec:discussion} and  \ref{sec:conclusion}.

\section{Method}
\label{sec:method}
\paragraph{}
We introduce the disk model in Section \ref{sec:diskmodel} and calculate the $r-$dependent planetesimal mass distributions in Section \ref{sec:local_dist_plts}.
The N-body numerical setups are described in Section \ref{sec:simul_setup}.

\subsection{Disk model}
\label{sec:diskmodel}
This study focuses on the  planetesimal growth around the solar-mass central star. We adopt the MMSN disk model \citep{Hayashi1981}. The disk temperature, gas surface density, gas disk aspect ratio, and pressure gradient prefactor are expressed as  
\begin{equation}
    T = T_0 \left(\dfrac{r}{1 \text{AU}}\right)^{-1/2},
\label{eqn:tem_profile}
\end{equation}
\begin{equation}
    \Sigma_{\text{gas}} = \Sigma_{\text{gas},0}\left(\dfrac{r}{1 \text{AU}}\right)^{-3/2},
\label{eqn:den_profile}
\end{equation}
\begin{equation}
    h_{\text{gas}} = h_{\text{gas}, 0}\left(\dfrac{r}{1 \text{AU}}\right)^{1/4},
\label{eqn:aspect_ratio_profile}
\end{equation}
and
\begin{equation}
    \eta {=} \eta_0 \left(\dfrac{r}{1 \text{AU}}\right)^{1/2},
\end{equation}
where $r$ is the disk radial distance, $\Sigma_{\text{gas},0} = 1700$ g cm$^{-2}$, $T_0 = 280$ K, $h_{\text{gas},0}= 0.033$, and $\eta_0 = 1.8\times10^{-3}$ are the corresponding values at $1$ AU. 

The detailed modeling of the dust growth and radial drift is not the focus of our study. Instead, we assume that the dust particles have already grown to pebble-sized bodies with a unified dimensionless stopping time (or termed Stokes number), and these pebbles sweep the disk from outside-in with a constant flux.  We chose the Stokes number of pebbles $\tau_{\rm s} {=} 0.1$ and the pebble flux of  $\dot{\text{M}}_{\text{peb}} {=} 100 \ \rm M_{\oplus}/Myr$ in the fiducial run. The above simplified treatment is a proper approximation from the more advanced dust coagulation models \citep{Birnstiel_etal2012, Lambrechts_Johansen2014,Drkazkowska2021}.

\subsection{Mass distribution of planetesimals in a ring}
\label{sec:local_dist_plts}
\paragraph{}
Since the disk properties are correlated with radial distance, the forming planetesimals are also expected to feature an $r$-dependence. We provide their mass distribution calculations as follows:  

\cite{Schafer_etal2017} showed that the initial mass function of planetesimals that form by streaming instability can be fitted by a power law plus an exponential decay function. The number fraction of planetesimals with $M_{\text{p}} > M$ is given by 
\begin{equation}
\dfrac{N_{>}(M)}{N_{\text{tot}}} = \left( \dfrac{M}{M_{\text{min}}} \right)^{-p}\text{exp}\left[ \left( \dfrac{M_{\text{min}}}{M_{\text{plt}}} \right)^{q} - \left( \dfrac{M}{M_{\text{plt}}} \right)^{q} \right],
\label{eq:Schafer2017}
\end{equation}
where $p {=} 0.6$ and $q{=}0.35$.
\Eq{Schafer2017} features two parameters: the minimum mass and the characteristic mass of planetesimals ($M_{\text{min}}$ and $M_{\rm plt}$). We find that the shape of the distribution and especially the planetesimal population that dominates the total mass are barely affected by $M_{\text{min}}$, as long as $M_{\text{min}}{\ll}M_{\rm plt}$ (see Appendix \ref{sec:appendix_minimum_mass_size_dist}). We thus set $M_{\rm min} {=} M_{\rm plt}/20$. Using a given characteristic mass $M_{\rm plt}$ and total mass of planetesimals in the belt $M_{\rm tot}$, we can obtain the planetesimal mass distribution from \Eq{Schafer2017}. As we show later, both $M_{\rm plt}$  and $M_{\rm tot}$ depend on the forming locations of the planetesimals.

We adopt the characteristic planetesimal mass from \cite{Liu_etal2020} based on the extrapolation of literature on streaming instability simulations, 
\begin{equation}
    \dfrac{M_{\text{plt}}}{M_{\oplus}} = 5\times 10^{-5} \left( \dfrac{Z}{0.02} \right)^{a} \left( \dfrac{\gamma}{\pi^{-1}} \right)^{a+1} \left( \dfrac{h_{\text{gas}}}{0.05} \right)^{3+b},
\label{eq:characteristic_mass0}
\end{equation}
where the fitting coefficients are $a {=} 0.5$ and $b{=}0$. The self-gravity parameter $\gamma$ is given by 
\begin{equation}
    \gamma = \dfrac{4\pi G \rho_{\text{gas}}}{\Omega_{\text{K}}^2},
    \label{eq:gamma}
\end{equation}
where the Keplerian orbital frequency $\Omega_{\text{K}}{\equiv} \sqrt{GM_{\star}/r^3}$, $G$ is the gravitational constant, $M_{\star}$ is the stellar mass, and $\rho_{\text{gas}}$ is the midplane gas density  
\begin{equation}
     \rho_{\text{gas}} = \dfrac{\Sigma_{\text{gas}}}{\sqrt{2\pi} H_{\text{gas}}},
\label{eq:rho}
\end{equation}
where $H_{\text{gas}}=h_{\rm gas} r$ is the gas disk scale height. 

In \Eq{characteristic_mass0}, the local metallicity $Z$ is estimated from the streaming instability triggering condition where $ \rho_{\text{peb}} {\simeq} \rho_{\text{gas}}$ \citep{Youdin_Goodman2005}. This can be equivalently expressed as $\Sigma_{\text{peb}}/H_{\text{peb}} \simeq \Sigma_{\text{gas}}/H_{\text{gas}}$. Hence, the threshold metallicity for the onset of streaming instability can be written as $Z_{\rm SI}{=}\Sigma_{\text{peb}}/\Sigma_{\text{gas}}{=}H_{\text{peb}}/H_{\text{gas}}$. The disk scale height ratio between pebbles and gas is given by \citep{Youdin_Lithwick2007}
\begin{equation}
    \dfrac{H_{\text{peb}}}{H_{\text{gas}}} = \sqrt{\dfrac{\alpha_t}{\alpha_t + \tau_s}},
     \label{eq:H_ratio}
\end{equation}
where $\alpha_t$ is the gas turbulent diffusion coefficient. We assume that the dust and gas have the same diffusivity for tightly coupled solid particles. Besides that, the coefficient of turbulent gas diffusivity approximates to the turbulent viscosity parameter for  magnetorotational instability (MRI) disks  \citep{Johansen_Klahr2005, Zhu_etal2015, Yang_etal2018}.  In this paper, we do not distinguish these two and adopt one unified parameter as $\alpha_t$. We set $\alpha_t{=}10^{-4}$ based on the observational diagnostic of disk turbulence \citep{Pinte_etal2016, Flaherty_etal2017}. Therefore, the threshold metallicity in the fiducial setup is $0.0316$, higher than the nominal solid-to-gas column density ratio of $0.01$.

Combining \Eqss{characteristic_mass0}{gamma}{rho}, we can rewrite the characteristic mass as 
\begin{equation}
  \dfrac{M_{\text{plt}}}{M_{\oplus}} = 1.2\times 10^{-6} \left(\dfrac{Z}{0.02}\right)^{1/2} \left(\dfrac{\Sigma_{\rm gas,0}}{1700}\right)^{3/2}
  \left(\dfrac{h_{\rm gas,0}}{0.033}\right)^{3/2} \left(\dfrac{r}{\text{3 AU}}\right)^{9/8}.
    \label{eq:characteristic_mass}
\end{equation}

The total mass of planetesimals in the ring belt can be estimated as 
\begin{equation}
    M_{\rm tot} = \epsilon M_{\rm tot, peb} = 2\pi \epsilon  r \Delta r \Sigma_{\text{peb}} = 2\pi \epsilon Z  r \Delta r \Sigma_{\text{gas}},
    \label{eq:total_mass}
\end{equation}
where $\epsilon$ is the pebble-to-planetesimal conversion efficiency. As explored and discussed in \cite{Abod_etal2019}, $\epsilon$ is at approximately ten percent and should vary with the radial pressure gradient, the Stokes number of pebbles, and the disk metallicity. However, due to limited numerical streaming instability simulation  explorations on $\epsilon$, we followed the same procedure as \cite{Liu_etal2019} and we set this efficiency at  a fixed value of $10\% $ for the sake of simplicity.
The width of the planetesimal belt can be estimated from the dense dust filament in streaming instability simulations, which is expressed as $\Delta r = \eta r$ \citep{Yang_Johansen2016,Li_etal2018}.

In sum, the planetesimal population can be calculated from \Eqss{Schafer2017}{characteristic_mass}{total_mass}, where the detailed numerical approach is illustrated in Appendix \ref{sec:appendix_mass_dist}. Table \ref{tab:characteristics_different_r}  summarizes the generated planetesimal population with its characteristic, minimum, and maximum masses in the MMSN model, while \Fg{size_dist} illustrates the discrete mass distributions at $r{=}0.3$ AU, $3$ AU, and $30$ AU, respectively. 

We note that streaming instability simulations produce numerous small planetesimals. From \Eq{Schafer2017}, the total number of planetesimals are dominated by low-mass bodies. Nevertheless, these bodies only contribute in a very limited way to the subsequent mass growth \citep{Liu_etal2019}. Conducting simulations with such a huge number of small planetesimals is numerically impractical. For the sake of computational feasibility, we simulate the  massive bodies (dark bars in \Fg{size_dist}), which corresponds to two-thirds of the total masses in the belt.

We also performed a simulation with $80\%$ of the total masses in the belt to investigate the effect of neglected masses on the simulations. Because of the difference in number and total mass of planetesimals, the final mass of the most massive planet is not exactly identical to the fiducial case. However, the overall features of the growth pattern (discussed in Section \ref{sec:results}) are very similar. It validates our approach in the main paper without exhaustively simulating all small planetesimals. 

\begin{table*}[]
\caption{Properties of planetesimals generated by streaming instability.}
\label{tab:characteristics_different_r}
\centering
\begin{tabular}{ccccccccc}
\noalign{\smallskip}
\hline\hline 
Description & $r$  {(}AU{)}  & $\tau_s$   & $N_{\text{tot}}$ & $N_{\text{sim}}$ & $M_{\text{plt}}$ {(}M$_{\oplus}${)} & $M_{\text{min}}$ {(}M$_{\oplus}${)} & $M_{\text{max}}$ {(}M$_{\oplus}${)} & $M_{\text{tot}}$ {(}M$_{\oplus}${)} \\ 
\hline
Fiducial & 0.3 & 0.1 &9943            & 924              & $1.3\times10^{-7}$                & $6.3\times10^{-9}$                & $1.3\times10^{-5}$                & $6.7\times10^{-4}$                \\ 
Fiducial & 3    &0.1 & 7470             & 651              & $1.7\times10^{-6}$                & $8.4\times10^{-8}$                & $1.5\times10^{-4}$                & $6.7\times 10^{-3}$               \\ 
Fiducial & 30  &0.1 & 5618             & 514              & $2.3\times10^{-5}$                & $1.1\times10^{-6}$                & $1.7\times10^{-3}$                & $6.7\times 10^{-2}$               \\ 
Low Stokes number & 0.3 & 0.03 & 13431 &1224 &$1.7\times 10^{-7}$ & $8.5\times10^{-9}$ &$1.9\times 10^{-5}$ & $1.2\times 10^{-3}$\\
\hline
\end{tabular}
\tablefoot{$N_{\text{tot}}$ is the total number of planetesimals in the belt, $N_{\text{sim}}$ is the number of selected planetesimals for numerical  simulations, $M_{\text{plt}}$ is the characteristic mass, and $M_{\text{tot}}$ is the total mass of the planetesimals. The Stokes number ($\tau_s$) of pebbles is $0.1$ in the fiducial run. For comparison, we set $\tau_s{=}0.03$ for the lower Stokes number run (Section \ref{sec:parameters}).}
\end{table*}

\begin{figure*}
    \centering
    \includegraphics[width=\textwidth]{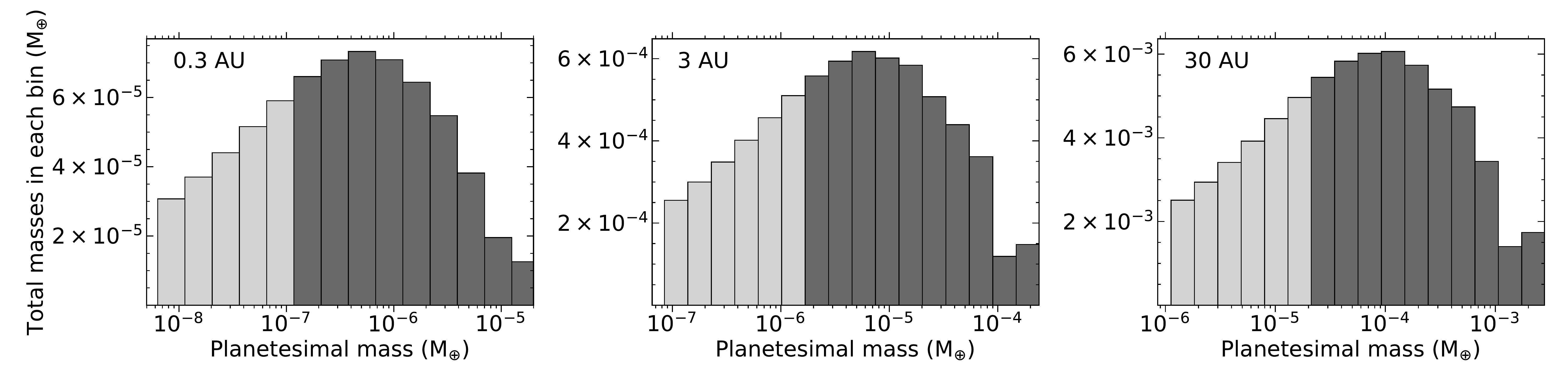}
    \caption{Initial mass distributions of planetesimals generated by streaming instability in a ring belt of the MMSN model at $r{=}0.3$ AU, $3$ AU, and $30$ AU, respectively. The darker bars represent the population of planetesimals with $M{>}M_{\rm plt}$, which account for two-thirds of the whole population in mass.  We only consider the above subset of planetesimals (darker) in the following numerical simulations. At 3 and 30 AU, the last darker bar is higher than the previous bar because the decimal places have been neglected when counting the number of planetesimals. If the number of planetesimals at the mass just less than the maximum mass is higher than one but less than two, it has been treated as one. Thus, the total mass of planetesimals in that bar becomes lower than their last bars.}
    \label{fig:size_dist}
\end{figure*}

\subsection{Numerical setup}
\label{sec:simul_setup}
\paragraph{}
The MERCURY N-body code \citep{Chambers1999} is adopted, where we choose the Burlirsch-Stoer integration, $3$ days for the initial timestep, and $10^{-12}$ for the integration accuracy. The code was modified to take into account of the effects of pebble accretion for planets/planetesimals, the gas drag for planetesimals, gravitational tidal torques for planets, and eccentricities/inclinations damping effects (see details in \citet{Liu_etal2019}).

Planetesimal accretion happens when the planetesimals collide with each other. For the N-body part, we simply assume that the collisions are ideally inelastic to conserve the angular momentum. Fragmentation is not taken into account, since the eccentricities of planetesimals are relatively small and the impact velocity is generally lower than the escape velocity. 

The pebble accretion prescription is adopted from \cite{Liu_Ormel2018} and \cite{Ormel_Liu2018}, where they account for the stochastic motion of pebbles due to disk turbulence. We also use the pebble isolation mass formula from \cite{Bitsch_etal2018}, where
\begin{equation}
    M_{\rm iso} = 25f_{\rm fit} \rm M_{\oplus},
\label{eq:Miso}
\end{equation}
where 
\begin{equation}
    f_{\rm fit} = \left[\dfrac{H/r}{0.05}\right]^{3}\left[0.34\left(\dfrac{\log(0.001)}{\log(\alpha_{\rm   t})}\right)^{4}+0.66\right]\left[1-\dfrac{\dfrac{\partial\ln P}{\partial \ln r}+2.5}{6}\right],
    \label{eq:f_Miso}
\end{equation}
$P$ is the gas pressure and $\alpha_{\rm t}$ stands for the turbulent viscosity parameter.

Once the planet reaches the pebble isolation mass, its mass growth via pebble accretion stops. Moreover, we simply assume that the other planets  located interior to this planet also terminate their pebble accretion. This is because a pressure bump is produced outside of the orbit of the planet that reaches the pebble isolation mass. The upcoming pebbles stop their radial drift exterior to this pressure bump and, as a result, all the inner bodies cannot further accrete pebbles. However, we note that a fraction of pebble fragments at the abovementioned pressure bump can still pass the gap and replenish the inner disk region \citep{Liu2022}. In addition, the pressure bump can also be the formation site of the next generation planetesimals, since the inward drifting pebbles are continuously accumulated \citep{Eriksson_etal2021, Eriksson_etal2022}. Such effects are not considered here. Additionally, we also do not take into account of the accretion of gas in this work. 

Planetesimals experience an aerodynamic gas drag, and its deceleration is expressed as 
\begin{equation}
    \boldsymbol{a_{\text{drag}}} = - \left(\dfrac{3C_{\text{D}}\rho_{\text{gas}}}{8R_{\text{p}}\rho_{\bullet}}\right) v_{\text{rel}}\boldsymbol{v_{\text{rel}}},
\end{equation}
where $C_{\text{D}}{=}0.5$ is the drag coefficient, $R_{\text{p}}$ is the radius of the planetesimal, $\rho_{\bullet}{=} 1.5 \rm \  g cm^{-3}$ is the internal density, and $\boldsymbol{v_{\text{rel}}}$ is the relative velocity between the planetesimal and gas $\boldsymbol{v_{\text{rel}}} {=} \boldsymbol{v_{\text{plt}}} - \boldsymbol{v_{\text{gas}}}$, where $\boldsymbol{v_{\text{gas}}}$ is the gas rotation velocity $\boldsymbol{v_{\text{K}}}(1-\eta)$, and $\boldsymbol{v_{\text{K}}}$ is the Keplerian velocity. 

When a planetesimal grows massive enough to excite density waves, these waves exert torques onto the planet and result in its orbital decay \citep{Kley_Nelson2012}. In our simulations, the highest mass that planets can reach is generally ${\lesssim}10 \ M_{\oplus}$  (see Sections \ref{sec:results} and \ref{sec:parameters}), and therefore only the type I torque is considered in our study. The corresponding accelerations are given by  
\begin{equation}
    \boldsymbol{a_m} = -\boldsymbol{v}/t_m, \  \boldsymbol{a_e} = -2\dfrac{(\boldsymbol{v \cdot r})\boldsymbol{r}}{r^2 t_e}, \ \boldsymbol{a_i} = \dfrac{\boldsymbol{v_z}}{t_i},
\label{eqn:a_mig}
\end{equation}
where $\boldsymbol{v}$ is the velocity vector of a migrating body, and $t_m$, $t_e$, and $t_i$ are the type I migration, eccentricity, and inclination damping timescales (see Equation (11-13) of \cite{Cresswell_Nelson2008}). The potential thermal radiative and dust feedback onto the type I migration torques is neglected in this work\citep{Jimenez_etal2017,Benitez-Llambay_Pessah2018}.

In order to minimize the number of free parameters and investigate the in situ growth, we set the semimajor axis damping to zero as $\boldsymbol{a_m}{=}0,$ while the eccentricity and inclination damping are included in Section \ref{sec:results} for the fiducial run. Other setups of simulations incorporating type I migration, and different values of pebble flux and Stokes number, are presented in Section \ref{sec:parameters}.

\section{Results}
\label{sec:results}
\begin{figure}
    \centering
    \includegraphics[width=\linewidth]{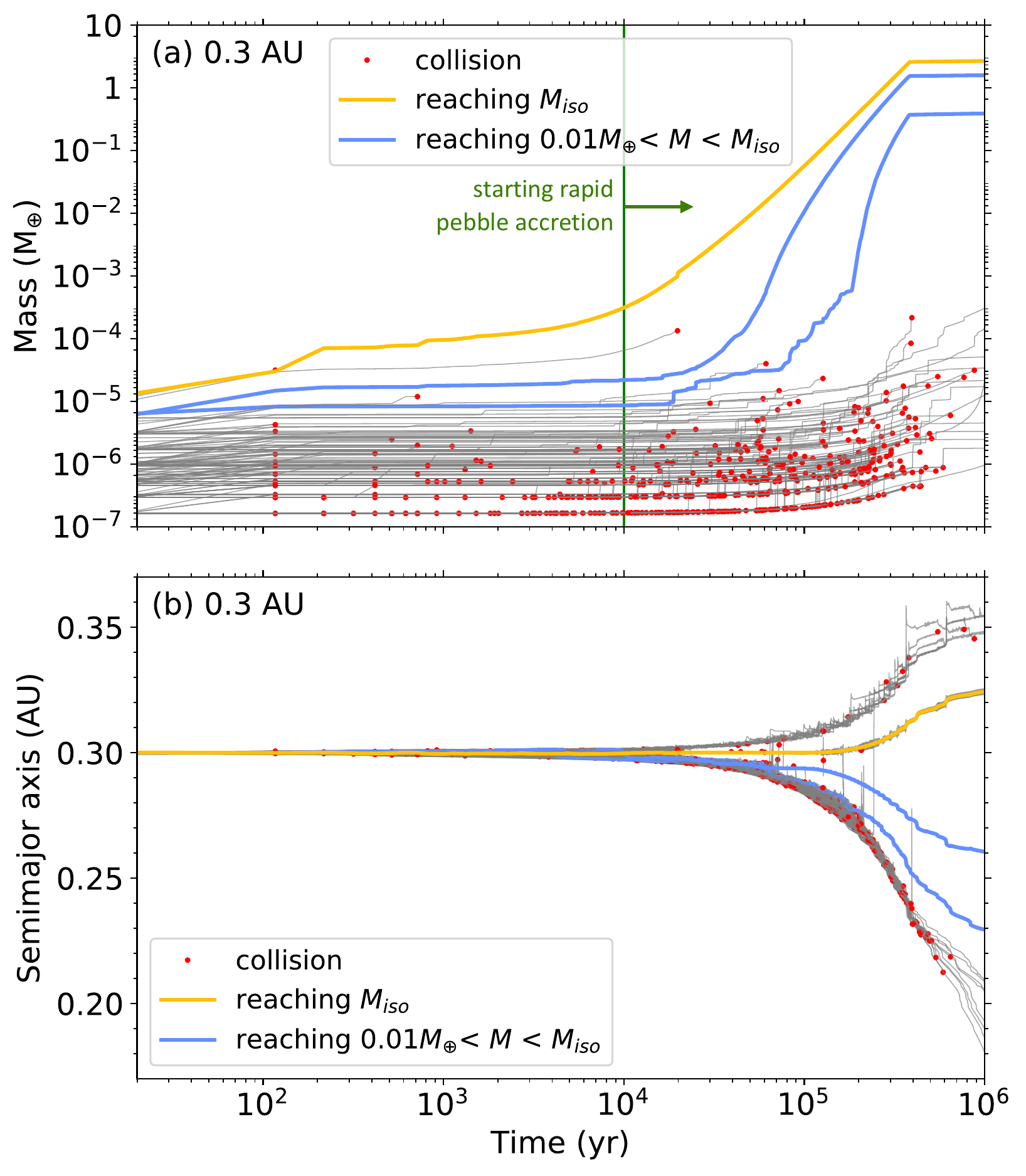}
    \caption{ (a) Mass growth and (b) semimajor  axis evolution of planetesimals in a ring located at $r{=}0.3$ AU. Each  line represents one planetesimal and red dots indicate planetesimal-planetesimal collisions. Gray lines represent all planetesimals in the simulation. The yellow and blue lines correspond to the bodies that finally reach the pebble isolation mass, and those that fail to reach the pebble isolation mass but grow above $10^{-2} \ M_{\oplus}$, respectively. The green horizontal line represents the time that large planetesimals start rapid pebble accretion. The mass growth is rapid at close-in orbits for both pebble and planetesimal accretion.}
    \label{fig:03simulation}
\end{figure}

\begin{figure}
    \centering
    \includegraphics[width=\linewidth]{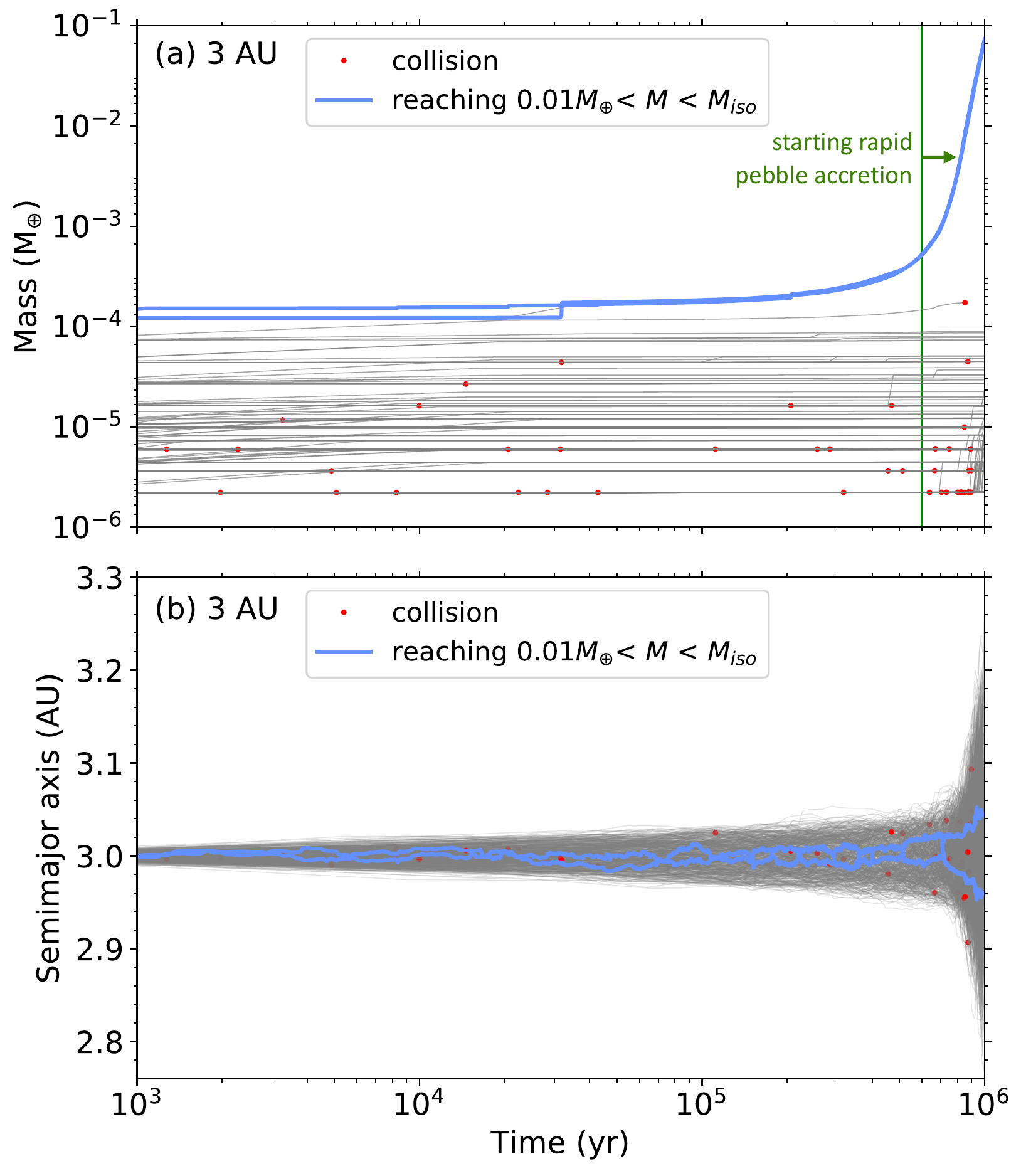}
    \caption{(a) Mass growth and (b) semimajor axis evolution of planetesimals in a ring located at $r{=}3$ AU. The same colored plots and lines are used as for \Fg{03simulation}. The dominant mass growth channel is pebble accretion.}
    \label{fig:3simulation}
\end{figure}

\begin{figure}
    \centering
    \includegraphics[width=\linewidth]{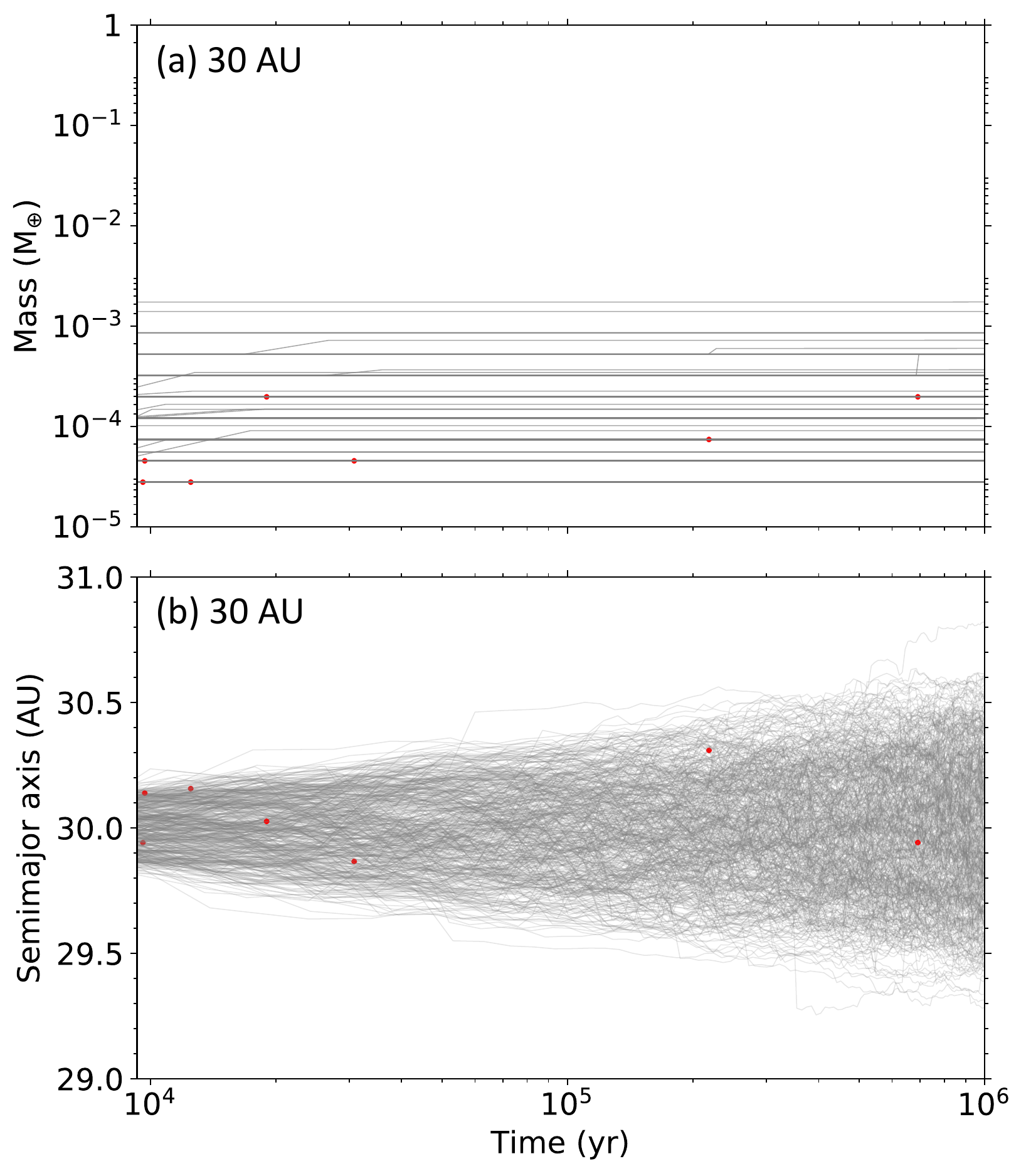}
    \caption{(a) Mass growth and (b) semimajor axis evolution of planetesimals in a ring located at $r{=}30$ AU. The notations in the graph are the same as \Fg{03simulation}. The growth of planetesimal is insignificant at such distant orbits in the MMSN disk model. }
    \label{fig:30simulation}
\end{figure}

\paragraph{}
In this section, we investigate the mass growth of planetesimals at three different disk locations. The model parameters are adopted from the fiducial run in Table  \ref{tab:parameter}.  Figure \ref{fig:03simulation}, \ref{fig:3simulation} and \ref{fig:30simulation} show the mass and semi-major axis evolution of the forming planetesimals at $r{=}0.3$ AU, $3$ AU, and $30$ AU for $1$ Myr. The gradual increase in mass is due to pebble accretion, while the sudden bumps are due to collisions among planetesimals (marked as red dots). These figures show planetesimals that do not undergo significant mass growth, that experience substantial mass growth but fail to reach the pebble isolation mass, and that  reach the pebble isolation mass within the simulated time, respectively. 


We see a fast growth of planetesimals when they form at $r{=}0.3$ AU in \Fg{03simulation}a. Three large planetesimals grow their masses effectively by pebble accretion, and one of them reaches the pebble isolation mass of $2.6 \ M_{\oplus}$ at $0.31$ AU (yellow line). The other two planetesimals stop their mass growth with $0.37 \ M_{\oplus}$ at $0.25$ AU and $1.5 \ M_{\oplus} $ at $0.27$ AU (blue lines). We find in \Fg{03simulation}b that the three massive growing embryos scatter each other and shield smaller planetesimals exterior and interior to their orbits after a few $10^5$ years.  

The growth is slower when the planetesimals are born at $r{=}3$ AU than at $0.3$ AU.  In \Fg{3simulation}a, after a few collisions, two initially massive planetesimals (${\sim}10^{-4} \ M_{\oplus}$) start rapid pebble accretion and finally reach around $ 0.07 \ M_{\oplus}$ (blue lines). The pebble isolation mass is $ 14\ M_{\oplus}$ at the corresponding radial distance. These two most massive planets stay around their initial semimajor axes, while other smaller planetesimals are continuously scattered by them into the exotic orbits (see \Fg{3simulation}b). Even though the planets do not reach the pebble isolation mass within 1 Myr, they reach the pebble isolation mass of $ 14\ M_{\oplus}$ within 2-3 Myr.

Although the generated bodies are more massive at $r{=}30$ AU compared to those at closer-in orbits (Table \ref{tab:characteristics_different_r}), we find that the mass growth is substantially suppressed when the planetesimals are born at a more distant disk location. The collisions are very rare in \Fg{30simulation}a and the planetesimals of $10^{-3} \ M_{\oplus}$ do not experience any mass increase. The width of the initial belt expands slowly, less than $50\%$ over $1$ Myr (see \Fg{30simulation}b). We will provide the analytical explanations for such $r$-dependent growth patterns in the following subsections. 

\subsection{Planetesimal accretion}
\label{sec:planetesimal_accretion}
\paragraph{}
The planetesimal collision rate can be expressed as
\begin{equation}
    P_{c}= n\sigma\delta v,
\label{eqn:col_rate1}
\end{equation} 
where $n$ is the number density of planetesimals expressed as $n = \Sigma_{\rm plt}/ M H$, where $M$ and $\Sigma_{\rm plt}$ are the mass and surface density of planetesimals, $\sigma$ is a collisional cross section area, and $\delta v$ is relative velocity. Since the vertical height of planetesimals $H \simeq \delta v/\Omega_{\rm K}$, we can rewrite Equation \ref{eqn:col_rate1} as
\begin{equation}
    P_{\rm c}=\Sigma_{\rm plt}\Omega_{\rm K} \sigma /M = \frac{3\Sigma_{\rm plt}\Omega_{\rm K}}{4 \rho_{\bullet} R_{\rm plt}} \left(1 +  \frac{v_{\rm esc}^2}{\delta v^2} \right),
\label{eq:col_rate2}
\end{equation} 
where $v_{\rm esc}$ and $R_{\rm plt}$ are the escape velocity and size of the planetesimals.  
We note that the mass of the planetesimal is lower, and the planetesimal surface density and the Keplerian orbital frequency are higher, at a smaller $r$. Even though the cross section $\sigma$ slightly decreases due to mutual excited orbits, the overall collision rate is much higher when $r$ is smaller. If we assume equal-sized planetesimals and the geometric accretion regime, the planetesimal collision timescale $\tau_{\rm c} {=}1/P_{\rm c} \propto r^{3}$. This is why, in the simulations, more frequent planetesimal-planetesimal collisions occur, and their mass growth  via planetesimal accretion is significant at close-in orbits. In total, there are $902$, $111$ and $22$ collisions for the cases at $r{=}0.3$, $3,$ and $30$ AU, respectively.

\subsection{Pebble accretion}
\label{sec:pebble_accretion}
\paragraph{}
The efficiency of pebble accretion correlates with the mass of the planetesimal/planet and the size of accreted pebbles \citep{Ormel_Liu2018}. Pebble accretion only becomes efficient when the planetesimal has a relatively high  mass; otherwise the gas-aided settling effect is insignificant  \citep{Visser_Ormel2016,Liu_Ji2020}. The onset mass for the efficient pebble accretion can be expressed as \citep{Liu_Ji2020}   
\begin{equation}
    M_{\text{onset}} =  \tau_s \eta^3 M_{\star} \simeq 10^{-3} \left(\frac{\tau_{\rm s}}{0.1}\right) \left( \frac{\eta_0}{1.8 \times 10^{-3}} \right)^3 \left(\dfrac{r}{\text{3 AU}}\right)^{3/2}  M_{\oplus}.
    \label{eq:m_crit}
\end{equation}

As shown in \Eq{m_crit}, the onset mass scales super-linearly with $r$ through the $\eta$ dependence. A more massive planetesimal is required to initiate the rapid pebble accretion at a larger orbital distance. For instance, $M_{\text{onset}} \simeq 3 \times 10^{-5} \ M_{\oplus}$ at $r{=}0.3$ AU, almost corresponding to the highest-mass planetesimal in the initially formed belt (Table \ref{tab:characteristics_different_r}). At $r{=}3$ AU, $M_{\rm onset}\simeq 10^{-3} \ M_{\oplus}$, a factor of $10$ higher than the highest-mass planetesimal in the belt. The mass difference is even larger at $r{=}30$ AU, where  $M_{\rm onset}\simeq 3 \times 10^{-2} \ M_{\oplus}$ and $M_{\rm max}{=}2\times10^{-3} \ M_{\oplus}$.
As a direct consequence, the planetesimals that form at a larger orbital distance need to grow more massive  in the first place, in order to trigger the efficient pebble accretion.

As discussed in Section \ref{sec:planetesimal_accretion}, the early mass increase is aided by planetesimal accretion, which also features a radial distance dependence. Significant mutual collisions among planetesimals occur faster at a smaller $r$. Therefore, the largest planetesimal reaches $M_{\rm onset}$ at an earlier time and consequently starts effective pebble accretion afterward. In \Fg{03simulation}a, the slope of the first growing planetesimal (yellow line) steepens at around $10^4$ years, while in \Fg{3simulation}a, the growing planetesimals steepen after $6\times10^5$ years. We do not see such a transition between two mass growth regimes in \Fg{30simulation}a.

The timescales of pebble accretion in 2D/3D regimes can be expressed as  \citep{Liu_etal2019} 
\begin{equation}
\begin{split}
    \tau_{\text{PA,2D}} \simeq &  6.5\times 10^4 \left(\dfrac{M}{0.05 \text{M}_{\oplus}}\right)^{1/3} \left(\dfrac{\tau_s}{0.1}\right)^{1/3}
    \left(\dfrac{\eta}{1.8\times10^{-3}}\right)\\ & \left(\dfrac{\dot{M}_{\text{peb}}}{100 \text{M}_{\oplus}/\text{Myr}}\right)^{-1} \text{yr}
\end{split}
\label{eq:time_peb2D}
\end{equation}
and
\begin{equation}
    \tau_{\text{PA,3D}} \simeq 1.5\times 10^4\left(\dfrac{h_{\text{peb}}}{1\times 10^{-3}}\right)\left(\dfrac{\eta}{1.8\times10^{-3}}\right)\left(\dfrac{\dot{M}_{\text{peb}}}{100 \text{M}_{\oplus}/\text{Myr}}\right)^{-1} \text{yr},
\label{eq:time_peb3D}
\end{equation}
where $h_{\rm peb}{=}H_{\rm peb}/r$  is the pebble disk aspect ratio (\Eq{H_ratio}). The above equations are valid when the planet mass is much higher than $ M_{\rm onset}$.
Since the pressure gradient prefactor $\eta$, the planetesimal mass, and the aspect ratio of pebble disk all increase with $r$, the adopted timescales are $\tau_{\text{PA,2D}} \propto r^{1/2}$ and  $\tau_{\text{PA,3D}} \propto r^{3/4}$.  Therefore, after reaching $M_{\rm onset}$, the planetesimals also take more time to grow their masses at a larger orbital distance through pebble accretion. 

Accounting for the above factors, we find that 1) although the initial birth mass and the total mass in the planetesimal belt are lower, the subsequent planetesimal growth pattern is still faster when they form at closer-in orbits, and 2) planetesimal accretion has a steeper $r$-dependence than pebble accretion, making it even more inefficient at larger orbital distances.    

\begin{table}[]
\caption{Simulation setups for the parameter study.  }
\label{tab:parameter}
\centering
\begin{tabular}{lcccc}
\hline\hline
Description                     & \begin{tabular}[c]{@{}c@{}}type I\\ migration\end{tabular}    &\begin{tabular}[c]{@{}c@{}}Pebble flux\\ {[}M$_{\oplus}$ Myr$^{-1}${]}\end{tabular} &  $\tau_{\rm s}$   \\
\hline
1. fiducial                            & No               & 100         & 0.1  \\
2. migration                           & Yes  & 100         & 0.1 \\
3. high pebble flux                    & No               & 200          & 0.1  \\
4. low Stokes number                          & No               & 100         & 0.03 \\
\hline
\end{tabular}
\tablefoot{Only one model parameter is varied for different cases compared to the fiducial setup. The semimajor axis decay due to type I migration is implemented in the migration case, the pebble flux is increased  to $200 \ M_{\oplus} \ \rm Myr^{-1}$ in the high pebble flux case, and  the Stokes number is reduced to $0.03$ in the low Stokes number case.}
\end{table}

\section {Parameter study}
\label{sec:parameters}
\paragraph{}
 In Section \ref{sec:results} we only conduct simulations with the fiducial setup. Here, we vary different model parameters to explore their influence on the planetesimal growth. Only one parameter is varied in each case compared to the fiducial case. For instance, we implement the type I migration and vary the pebble flux and the Stokes number of pebbles (see Table \ref{tab:parameter}). This parameter study is only performed at $r{=}0.3$ AU, since it is the disk location where the planetesimals actively grow. The parameter study at 2.7 AU, similar to 3 AU, is performed in \cite{Liu_etal2019}.
 In order to account for the statistical variations, we run three different realizations for each case by randomizing the initial locations and phase angles of forming planetesimals. The resultant planets  that are more massive than $0.01 \ M_{\oplus}$ are shown in \Fg{param_study}. 
 
\begin{figure}
    \centering
    \includegraphics[width=\linewidth]{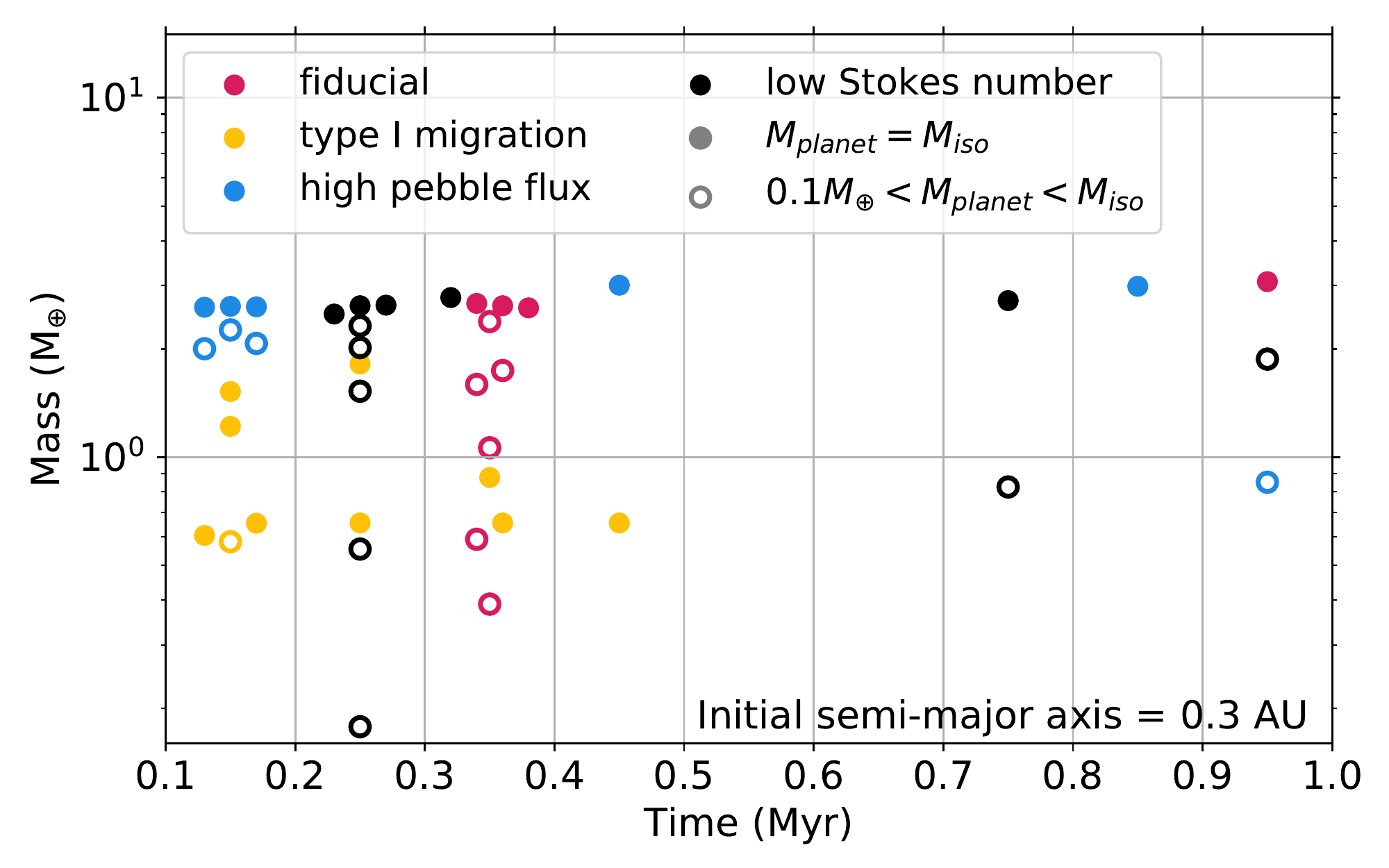}
    \caption{Final planet masses vs. time when they reach these masses around $r=0.3$ AU of the MMSN model. The planets with masses larger than 0.1 M$_{\oplus}$ are presented, including the fiducial case (red), the type I migration case (yellow), the high pebble flux case (blue), and the low Stokes number case (black). The filled circles are bodies that reach their pebble isolation masses within 1 Myr, while unfilled circles are bodies that experience effective mass growth above $0.1$ M$_{\oplus}$ but fail to reach the pebble isolation mass. Time is plotted in ranges, such as 0.1-0.2 Myrs or 0.2-0.3 Myrs. When the plots are overlapped, they are replaced horizontally on the side in the same time range.}
    \label{fig:param_study}
\end{figure}

First, we include type I migration to see how this process affects the growth of planetesimals. We assume that planets stop their migration at $r{=}0.05$ AU because of the gas clearing due to the central stellar magnetic torque \citep{Liu_etal2017}. In the migration case, the embryos migrate out of the birth planetesimal belt once their masses grow beyond $0.01 \ M_{\oplus}$ (\Fg{03mig}). The random velocities of the remaining planetesimals are less excited afterward, and these low-mass bodies can grow faster. The migrating embryos grow their masses by pebble accretion and quickly reach the inner disk edge. Since the pebble isolation mass is lower at a smaller radial distance, the final planets in \Fg{param_study} tend to have lower masses in the migration case (yellow) compared to those in the fiducial case (red). Those embryos can collide with each other actively through giant impacts (\Fg{03mig}b).

\begin{figure}
    \centering
    \includegraphics[width=\linewidth]{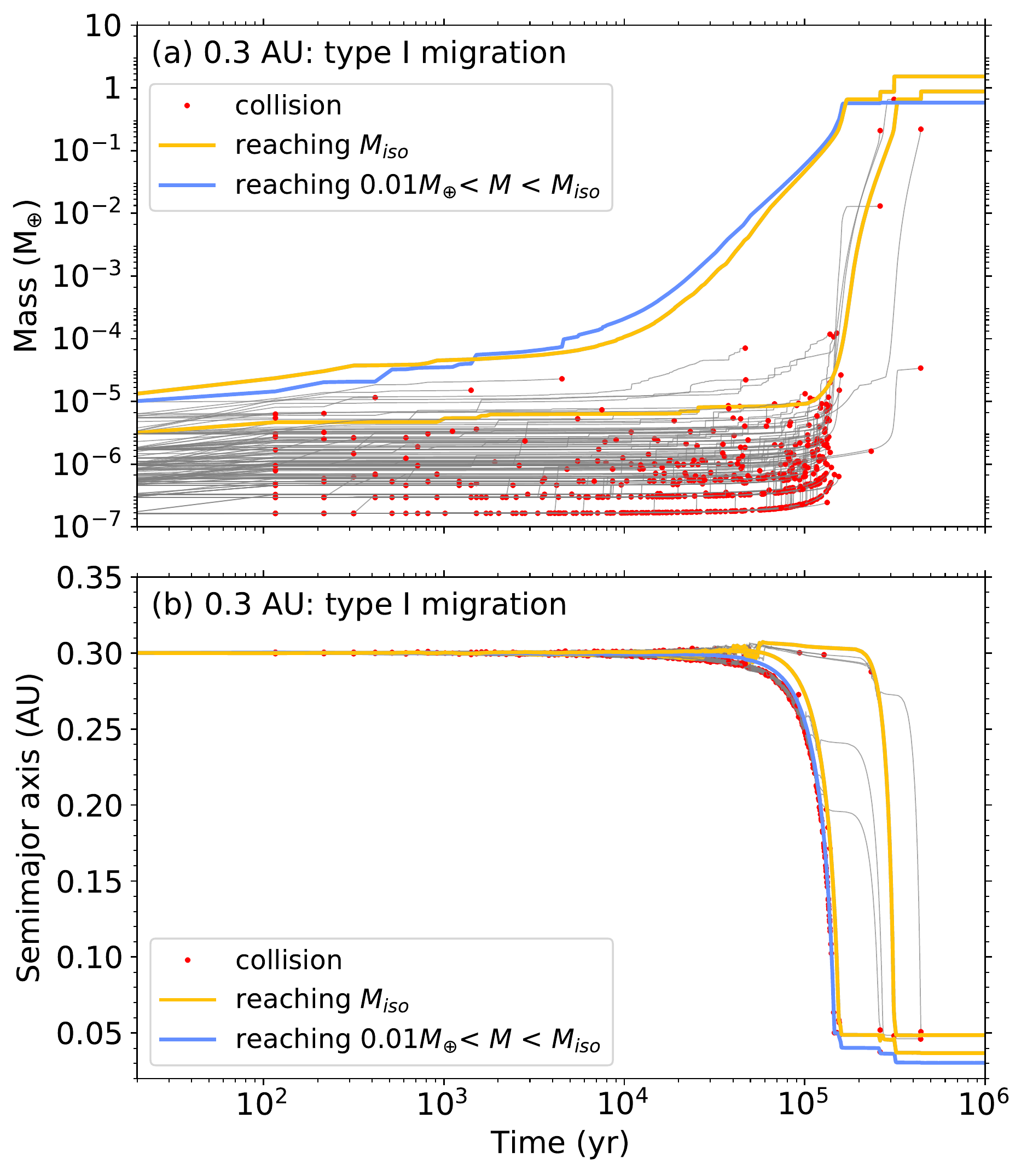}
    \caption{(a) Mass growth and (b) semimajor axis evolution of planetesimals in a ring located at $r{=}0.3$ AU with type I migration. The same colored plots and lines are used as for \Fg{03simulation}. The  embryos migrate out of the birth ring when their masses exceed $10^{-3} \  M_{\oplus}$ and stop pebble accretion at the inner edge of the disk. They can further increase their masses via giant impacts.}
    \label{fig:03mig}
\end{figure}

The pebble flux is a representative of the available building blocks for planet growth. We increase the pebble flux to $200$ M$_{\oplus}$ Myr$^{-1}$ in the high pebble flux case. We find in \Fg{param_study} that Earth-mass planets reach the pebble isolation mass at an earlier time  in this case (blue), and their masses are similar  to those formed in the fiducial case. For instance, in the fiducial case, the planets reach the pebble isolation mass between $t{\sim}0.3{-}0.4$ Myr, while in the high pebble flux case, the planets reach the pebble isolation mass as early as $t{\sim}0.1{-}0.2$ Myr. This can be understood from \Eq{time_peb2D} and \Eq{time_peb3D}, since the growth timescale is inversely proportional to the pebble flux.

We chose $\tau_{\rm s}{=}0.03$ in the low Stokes number case. In \Fg{param_study}, planets also reach the pebble isolation mass at earlier times (black) of $t{\sim}0.2{-}0.3$ Myr. The reasons are explained as follows. First, since the threshold metallicity for triggering streaming instability $Z_{\rm SI}$ depends on the Stokes number, a lower $\tau_{\rm s}$ leads to a higher metallicity through \Eq{H_ratio} and, therefore, more massive birth planetesimals (see Table \ref{tab:characteristics_different_r}). Second, the onset mass for pebble accretion is also lower in the low Stokes number case. In \Eq{m_crit}, the onset mass with $\tau_s = 0.03$ at $0.3$ AU is $M_{\rm onset}\simeq 9.5\times 10^{-6}$, which is similar to $M_{\rm max} = 1.9\times 10^{-5}$ (the maximum body from the birth population). Thus, large planetesimals exceed the onset mass threshold earlier for rapid pebble accretion in the low Stokes number case.

\section{Planetesimal growth in the Toomre limit}
\label{sec:superEarth}
\paragraph{}

The Atacama Large Millimeter/submillimeter Array  (ALMA) has revealed that rings and gaps are common in young protoplanetary disks at the orbital distances of a few tens to hundreds of AUs \citep{Avenhaus_etal2018, Huang_etal2018,Long_etal2019,Cieza_etal2019}. One leading explanation is that these substructures are caused by embedded planets due to planet-disk interactions (but also see \citet{Jiang2021}). This explanation requires that the corresponding planetary mass is around Neptunian to Jovian mass \citep{Zhang_etal2018,Bae_etal2018,LiuSF2018, Lodato_etal2019}. Here, we also attempt to explore whether massive protoplanets can form at young ages of $1{-}3$ Myr from our described configuration, where the streaming instability is triggered at a single disk location and a ring of planetesimals form. In the very early disk phase, the disk surface density can be much higher than in the fiducial MMSN model.  

We estimate the upper limit of the gas surface density for a stable disk from the Toomre criterion, where the Toomre parameter is given by \citep{Toomre1964} 
\begin{equation}
    Q \equiv \dfrac{c_s \Omega_{\rm K}}{\pi G \Sigma_{\text{gas}}} = h_{\rm gas}\frac{M_{\star}}{\pi r^2 \Sigma_{\rm gas}},
\label{eq:toomre}
\end{equation}
where $c_s{=}H_{\rm gas}\Omega_{\rm K}$ is the gaseous sound speed, and $Q{<} 1$ corresponds to an unstable disk. Thus, the upper limit of the gas surface density for a stable disk can be written as
\begin{equation}
   \Sigma_{\text{gas, Q}} = 240 \left(  \frac{h_{\rm gas, 0}}{0.033}\right) \left(\dfrac{r}{30 \rm \ AU}\right)^{-7/4} \text{g cm$^{-2}$}.
\end{equation}

The Toomre gas surface densities are 240 g/cm$^2$ and 30 g/cm$^2$ at $30$ AU and $100$ AU, which are $24$ and $18$ times higher than the corresponding values in the MMSN disk model. 
The planetesimals could have the highest masses if streaming instability is triggered at such a high threshold disk density condition.  We  test planetesimal formation in this extreme regime.  The properties of the formed planetesimals are summarized in Table \ref{tab:Toomre_gas}. Since the disk density is higher at such an early stage, we also explore cases with a different pebble flux of $100$, $200,$ and $300 \ \rm M_{\oplus} \ Myr^{-1}$. We note that the mass of  the planetesimal in this case can be treated as an upper limit, and more realistic final masses of the planets should likely lie somewhere between the Toomre and MMSN circumstances.

We performed simulations for two different cases: either with a planetesimal mass distribution based on \Eq{Schafer2017} or with only the single largest planetesimal from that distribution. In the former case, both the N-body interaction and pebble accretion are taken into account, while in the latter case, only pebble accretion is considered. We compare these two cases at $t{=} 1$ Myr. \Fg{QSig} shows the masses of the embryos that grow  by $50\%$ from their initial masses and are greater than 0.1 M$_{\oplus}$ at the end of the simulations. 

We find, in \Fg{QSig}, that the planetesimal-planetesimal interaction has little contribution to the mass growth of the most massive body. This is in line with our expectation that pebble accretion is more predominant for the subsequent planetesimal growth at the more distant disk location (Section \ref{sec:results}). We continue to simulate the growth of the single planetesimal for another $2$ Myr. Their masses at $t{=}2$ Myr and $3$ Myr are plotted in \Fg{QSig} with larger unfilled circles. For each case, the empty plots are denoted next to each other, but they all have the same semimajor axes of 30 AU or 100 AU. For comparison, the gap opening masses based on Eq. (35) of \cite{Liu_etal2019Dec}, followed by \cite{Kanagawa_etal2015}, are also depicted with horizontal dotted lines. With the explored parameters in the Toomre's disk, we find that the forming planets can have highest masses up to $54 \ M_{\oplus}$ at $r{=}30$ AU, which exceeds the gap opening mass, and up to $24 \ M_{\oplus}$ at $r{=}100$ AU within $3$ Myr. In a low turbulent disk of $\alpha_{\rm t} {=} 10^{-4}$, the pebble isolation mass obtained from \cite{Bitsch_etal2018} is higher than the gap opening mass calculated from \cite{Kanagawa_etal2015}. For  illustrative purposes, we continue our simulations in \Fg{QSig}, even when the planets reach the gap opening mass.

\begin{table*}[]
\caption{ Properties of planetesimals generated by streaming instability at $r{=} 3$ and $30$ AU in an upper limit of  the stable disk under the Toomre's criterion.}
\label{tab:Toomre_gas}
\centering
\begin{tabular}{ccccccc}
\noalign{\smallskip}
\hline\hline 
Radial distance & $N_{\text{tot}}$ & $N_{\text{sim}}$ & $M_{\text{plt}}$ {[}M$_{\oplus}${]} & $M_{\text{min}}$ {[}M$_{\oplus}${]} & $M_{\text{max}}$ {[}M$_{\oplus}${]} & $M_{\text{tot}}$ {[}M$_{\oplus}${]} \\
\hline
30 AU           & 2225             & 199              & $2.6\times10^{-3}$               & $1.3\times10^{-4}$               & $1.29\times10^{-1}$                & $1.7$                             \\
100 AU          & 3096             & 273              & $6.4\times10^{-3}$               & $3.2\times10^{-4}$               & $3.7\times10^{-1}$                & $4.2$    \\ 
\hline
\end{tabular}
\end{table*}

\begin{figure*}
    \centering
    \includegraphics[width=\textwidth]{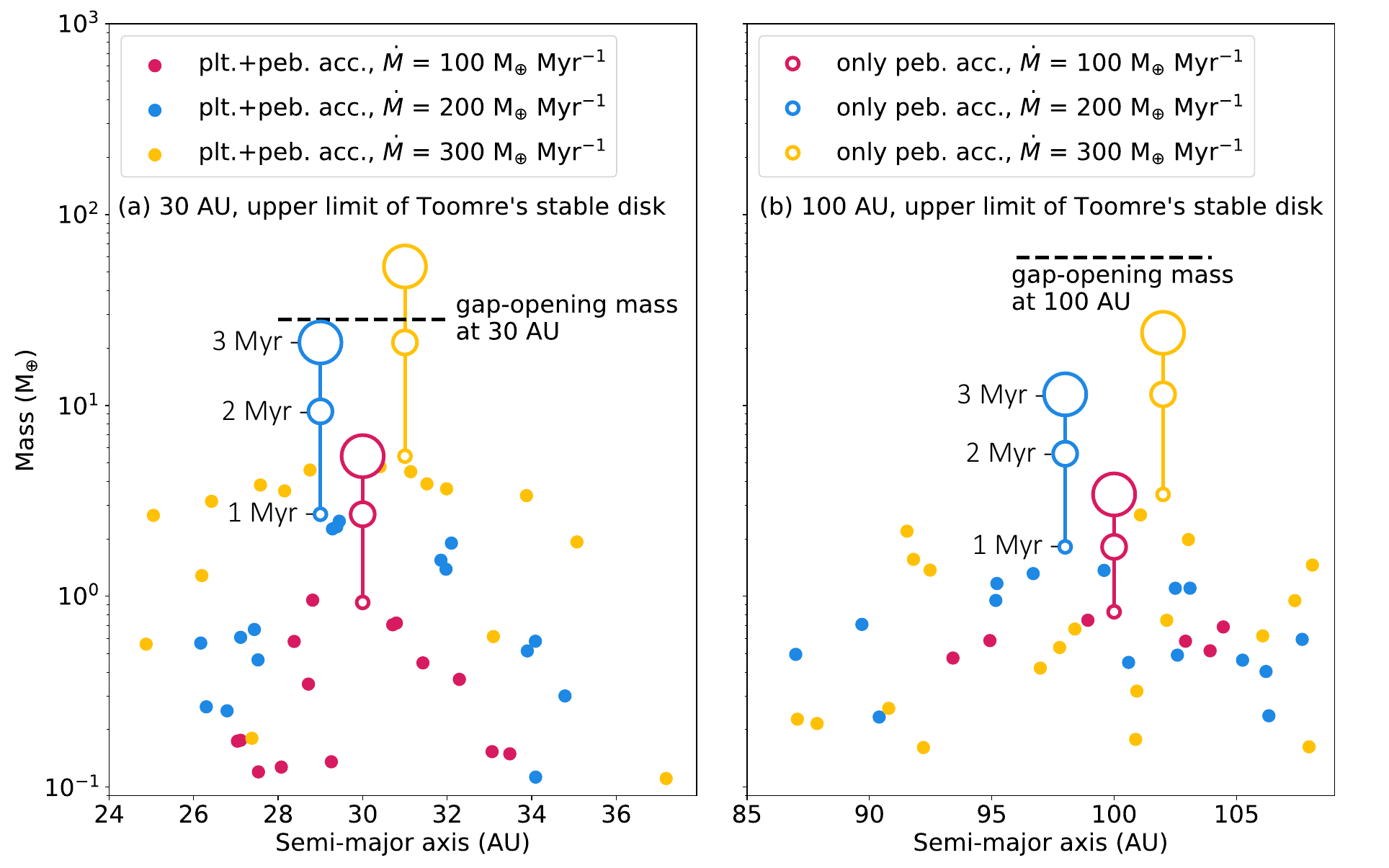}
    \caption{Final masses and semimajor axes of formed planets from planetesimal rings at (a) $30$ AU and (b) $100$ AU orbital distances based on Toomre's surface density criterion. The filled dots represent the final bodies from the planetesimal population (\eq{Schafer2017}) that grow by $50\%$ from their initial masses and are greater than $0.1 \rm M_{\oplus}$ at $t{=}1$ Myr. The unfilled circles denote  the single largest planetesimal that only undergoes pebble accretion, for comparison. The three unfilled circles from small to large refer to the mass at $t{=}1$, $2,$ and $3$ Myr, respectively. Red, blue, and yellow correspond to the pebble flux of $100$, $200,$ and $300 \ \rm M_{\oplus} \ Myr^{-1}$, respectively. In order to distinguish the mass more clearly, we artificially shift the blue and yellow unfilled circles slightly away from the birth locations. The black dotted horizontal lines represent the gap-opening masses based on Eq. (35) of \cite{Liu_etal2019Dec} at $30$ AU and $100$ AU.}
    \label{fig:QSig}
\end{figure*}

\section{Discussion}
\label{sec:discussion}
\paragraph{}
The disk turbulence plays a crucial role in regulating the rate of planetesimal formation via streaming instability. Many theoretical studies investigate how the growth of the streaming instability is influenced by disk turbulence \citep{Umurhan_etal2020,Gole_etal2020,Chen_etal2020,Schafer_etal2020,Xu_Bai2022}. For instance, \cite{Chen_etal2020} find that, since the dust particles get stirred by turbulence,  the growth rate of streaming instability is largely reduced. This suppression becomes insignificant when $\alpha_{\rm t}$ is low and $\tau_{\rm s}$ is high. On the other hand, \cite{Schafer_etal2020} find that the streaming instability becomes more pronounced when it coexists with the  vertical shear instability. A similar positive effect is also confirmed  by \cite{Xu_Bai2022} in their magnetohydrodynamic simulations. In these circumstances, depending on the nature of turbulence, the disk exhibits local, temporary pressure bumps, promoting the concentration of dust particles into clumps. Nevertheless, there is a lack of systematic work on how the disk turbulence affects the birth mass of forming planetesimals. Although our study assumes a weakly turbulent disk of $\alpha_{\rm t}$, the initial planetesimal mass function is adopted from \cite{Schafer_etal2017}, where turbulence is not accounted for in their simulations.

The formation of massive giant planets at distant orbits
 has   also been studied in other recent work \citep{Rosenthal_Murray-Clay2018,Najita_etal2021,Chambers2021}. For instance, \cite{Najita_etal2021} consider various initial disk solid mass and planetesimal formation efficiency and find that  Pluto-, Mars-, or super-Earth-mass planets form at greater than $30{-}40$ AU in their evolutionary models.  \cite{Chambers2021} studies multiple pressure bumps in the inner and outer regions of the disk with different birth masses of the planetesimals and disk turbulent strength. The author finds that a Jupiter-mass planet can form at $60$ AU within $0.5 {-}3$ Myr in a $0.03 \ M_{\odot}$ disk. \cite{Rosenthal_Murray-Clay2018} also study the formation of a gas giant at a wide orbit via a combination of planetesimal and pebble accretions with different levels of turbulence of the disk. The authors find that a gas giant can form at $\lesssim$ 40AU in a strongly turbulent disk and at $\lesssim$ 70AU in a weakly turbulent disk. 
 
Similarly, we also tested whether massive planets could form in the  outer disk region of $30$ and $100$ AU. 
We consider the upper limit of a stable disk with Toorme's criterion and tested different pebble fluxes. We find that several super-Earth planets form within $1$ Myr when $100{-}300 \ M_{\oplus}$ pebbles drift across the orbit of the planetesimal ring. The giant planets of $20{-}60 \ M_{\oplus}$ form after $3$ Myr at the corresponding disk radii when $900 \ M_{\oplus}$ pebbles sweep by. 

Noticeably, \cite{Najita_etal2021} and \cite{Chambers2021} investigate planet formation with pressure bumps. They infer the  total mass of forming planetesimals from dust mass in the observed rings. Our study is, however, not limited to this circumstance. Specifically, we consider the situation when  streaming instability is triggered at a single disk location. This is irrelevant with the presence of pressure bumps. Meanwhile, in our work, the total mass of planetesimals is  calculated according to the streaming instability triggering condition \citep{Youdin_Goodman2005}.

In addition, the starting mass of the planetesimal is the key free parameter in other studies \citep{Chambers2021}. We self-consistently calculate the planetesimal mass distribution based on realistic streaming instability simulations \citep{Schafer_etal2017,Liu_etal2020}. The masses of birth planetesimals vary with disk radial distances.    We fixed the turbulent strength at $10^{-4}$. This corresponds to the optimized configuration of giant planet  formation in \cite{Chambers2021}.

\cite{Rosenthal_Murray-Clay2018} suggest that the formation of gas giants at wide orbits is difficult, due to inefficient planetesimal and pebble accretion; but pebble accretion is still powerful enough to form smaller planets. This is in line with our results. As pebble accretion is still effective at forming a few Earth-mass planets in \cite{Rosenthal_Murray-Clay2018}, we also find that super-Earth planets naturally form at $30$ and $100$ AU in a quiescent and denser disk. Although \cite{Rosenthal_Murray-Clay2018} do not find that massive cores  trigger gas accretion, we find tens of Earth-mass planets with a high pebble flux at the upper limit of Toomre's stable disk, within a growth time of $3$ Myr.

\section{Conclusion}
\label{sec:conclusion}
\paragraph{}

In this work, we investigate the planetesimal growth by a combination of pebble and planetesimal accretion after their formation via the streaming instability mechanism at a single disk location. A ring belt of planetesimals is assumed to form in such a configuration \citep{Liu_etal2019}.  We particularly focus on how the growth pattern depends on radial distance $r$. 
The planetesimals with initial mass distributions based on the streaming instability simulations are generated at different disk radii of $0.3$, $3$, $30,$ and $100$ AU. The characteristic mass of the birth planetesimals increases with orbital distance and gas surface density \citep{Liu_etal2020}.  We simulate their subsequent mass growth and dynamical evolution. The key conclusions are listed as follows:
\begin{itemize}

\item[$\bullet$] We find that both planetesimal and pebble accretion are more active at smaller orbital distances. Although they both become slower at more distant disk locations, the radial distance dependence of pebble accretion is nevertheless weaker than that of planetesimal accretion (Section \ref{sec:results}).  

\item[$\bullet$] In the fiducial MMSN disk model, 
 several planets with masses of ${\sim}0.1{-}3 \ M_{\oplus}$  emerge after $1$ Myr at the close-in orbit of $0.3$ AU (\Fg{03simulation}).  The planets have lower final masses at larger orbital distances, with ${\sim}0.1 \ M_{\oplus}$ at $3$ AU and ${\sim}10^{-3} \  M_{\oplus}$ at $30$ AU, respectively (\Fgs{3simulation}{30simulation}). The planetesimal growth is strongly limited at  $r{=}30$ AU for the MMSN model.

\item[$\bullet$] The influence of different model parameters on the growth pattern is explored in Section \ref{sec:parameters}. Compared to the fiducial model, we find that when planetary orbital migration is considered, a greater number of Earth-mass planets form earlier around the inner edge of the disk (\Fg{03mig}).

\item[$\bullet$] The planets reach their pebble isolation masses at earlier times when the disks have a higher pebble flux or a lower Stokes number of pebbles (\Fg{param_study}).  

\item[$\bullet$] In the stable disks with Toomre's gas surface density at $30$ and $100$ AU (Section \ref{sec:superEarth}), planets can approach the gap opening mass within $2{-}3$ Myr after their formation, depending on the pebble flux. The final planet can attain $54 \ M_{\oplus}$ at $30$ AU and $24 \ M_{\oplus}$ at $100$ AU within $3$ Myr  by a given pebble flux of $300 \ M_{\oplus} \  \rm   Myr^{-1}$ (\Fg{QSig}).  Since the generated planetesimals are massive enough to trigger rapid pebble accretion in this high density disk condition, the growth is significantly dominated by efficient pebble accretion.
\end{itemize}

In this paper, we focus on the radial distance dependence. In future work, we will extend the study to look at the growth of planetesimals after the streaming instability around different mass stars. 

\section*{Acknowledgments}
We thank the anonymous referee for their insightful comments. H. J. acknowledges the finical support from A.J. for the summer student internship, which simulates the idea of this work. B. L. is supported by the start-up grant of the Bairen program from Zhejiang University, National Natural Science Foundation of China (No.12173035 and 12111530175), the Fundamental Research Funds for the Central Universities (2022-KYY-506107-0001). A.J. acknowledges funding from the European Research Foundation (ERC Consolidator Grant 724687-PLANETESYS), the Knut and Alice Wallenberg Foundation (Wallenberg Scholar Grant 2019.0442), the Swedish Research Council (Project Grant 2018-04867), the Danish National Research Foundation (DNRF Chair Grant DNRF159) and the G\"oran Gustafsson Foundation.  

\bibliographystyle{aa}
\bibliography{ref}

\begin{appendix}
\section{Variation of $M_{\rm min}$ in the planetesimal mass distribution}
\label{sec:appendix_minimum_mass_size_dist}
\paragraph{}
In the main paper, $M_{\rm min}$ generated from the streaming instability is set to be $20$ times smaller than the characteristic mass $M_{\rm plt}$. We note that this factor (i.e., how low the mass of the planetesimal can be) is actually less well constrained. For instance, the smallest size of the planetesimal is largely limited by the grid resolution in the streaming instability numerical simulations \citep{Schafer_etal2017}. Here, we vary different values of $M_{\rm min}$ and re-generate the planetesimal populations by a fixed total mass of planetesimal in the belt. The results of $M_{\rm min}{=}0.01 M_{\rm plt}$ are listed in Table \ref{tab:lower_min_mass}.  When comparing these populations with those presented in  Table \ref{tab:characteristics_different_r}, we find that the highest mass of the planetesimal, as well as the mass distribution, are barely affected by the variation of $M_{\rm min}$. This is valid as long as $M_{\rm min}{\ll}M_{\rm plt}$.

\begin{table}[h]
\caption{Planetesimal population based on \Eq{Schafer2017} by adopting $M_{\rm min}{=}0.01 M_{\rm plt}$. }
\label{tab:lower_min_mass}
\centering
\begin{tabular}{ccccccc}
\noalign{\smallskip}
\hline\hline 
$r$ & $M_{\text{min}}$ {[}M$_{\oplus}${]} & $M_{\text{max}}$ {[}M$_{\oplus}${]}  \\ 
\hline
0.3 AU          & $1.3\times 10^{-9}$              & $1.2\times 10^{-5}$                            \\
3 AU            & $1.7\times 10^{-8}$              & $1.4 \times 10^{-4}$                           \\
30 AU           & $2.3\times 10^{-7}$              & $1.7\times 10^{-3}$                 \\
\hline
\end{tabular}
\tablefoot{The mass of the maximum body and the shape of the mass distribution are still the same compared to the cases ($M_{\rm min}{=}0.05 M_{\rm plt}$)  explored in the main text (see Table \ref{tab:characteristics_different_r}). }
\end{table}

\section{Calculation of the mass distribution of planetesimals}
\label{sec:appendix_mass_dist}
\paragraph{}
\Eq{Schafer2017} is a cumulative number fraction of planetesimals, so we converted the fractions to be noncumulative $N(m)/N_{\text{tot}}$ by subtracting next number fraction bin. Then, we were able to obtain the number of planetesimals from the noncumulative number fractions by multiplying the total number of planetesimals. The total number of planetesimals is estimated from the sum of all the mass fractions for all the mass bins, which is represented by $M_{\text{tot}}/N_{\text{tot}}$. Therefore, the total number of planetesimals is 
\begin{equation}
    N_{\text{tot}} = \dfrac{M_{\text{tot}}}{M_{\text{tot}}/N_{\text{tot}}} = \dfrac{M_{\text{tot}}}{\sum\dfrac{N(m)}{N_{\text{tot}}}\times m}.
\end{equation}

With this total number, we were able to convert the number fraction ($N/N_{\text{tot}}$) of planetesimals to the number of planetesimals for each mass bin. Then, we determined the maximum mass of the local distribution to be the mass that contains only one planetesimal. Therefore, we have the local mass distributions from the minimum mass to the maximum mass. Since the simulations only take integer number of planetesimals, we ignored the numbers in decimal places. 
\end{appendix}
\end{document}